\documentclass[nohyperref]{article}
\usepackage[numbers, sort, compress]{natbib}
\usepackage[margin=.9in]{geometry}
\usepackage[dvipsnames]{xcolor}
\usepackage[utf8]{inputenc} 
\usepackage[T1]{fontenc}    
\usepackage{hyperref}       
\usepackage{booktabs}       
\usepackage{amsfonts}       
\usepackage{nicefrac}       
\usepackage{listings}
\usepackage{microtype}      
\usepackage{mathtools}
\usepackage{refcount}
\usepackage{algorithm}

\usepackage{algpseudocode}
\usepackage{amsmath,amssymb, amsthm}
\usepackage{physics}
\usepackage{caption}
\usepackage[bottom]{footmisc}
\usepackage{subcaption}
\usepackage{authblk}

\usepackage{times}  
\usepackage{helvet}  
\usepackage{courier}  
\usepackage{algorithm}
\usepackage{algpseudocode}
\usepackage{anyfontsize}
\usepackage{newfloat}
\usepackage{listings}
\DeclareCaptionStyle{ruled}{labelfont=normalfont,labelsep=colon,strut=off} 
\floatstyle{ruled}
\newfloat{listing}{tb}{lst}{}

\usepackage[utf8]{inputenc}
\usepackage{authblk}

\usepackage{bibentry}

\usepackage{footnote}
\usepackage{tablefootnote}
\usepackage{enumerate}
\usepackage{amsthm}
\usepackage{amsmath}
\usepackage{amssymb}
\usepackage{amsfonts}

\usepackage{esvect}
\usepackage{appendix}
\usepackage{import}
\usepackage{subcaption}
\usepackage{enumitem}
\usepackage{booktabs}
\usepackage{cleveref}
\usepackage{tikz}
\usepackage{graphicx}
\usetikzlibrary{shapes.geometric, arrows}
\usepackage{nameref}

\newcommand{\n}{n}                   

\newcommand{\CandidatesNum}{c}       
\newcommand{\Candidates}{\mathcal{C}}
\newcommand{\cand}{C}                
\newcommand{\candset}{X}             
\newcommand{\candsetsize}{k}         

\newcommand{\ranking}{\sigma}        
\newcommand{\ranki}[1]{{\sigma[#1]}}   

\newcommand{\SlotsNum}{s}            
\newcommand{\Slots}{\mathcal{S}}
\newcommand{\slot}{S}

\newcommand{\Rel}{R}                
\newcommand{\RelSamples}{\mathcal{R}}

\newcommand{\PR}{\hat{P}(R)}        

\newcommand{\mbm}[3]{{\mbox{MBM}(#1,#2|#3)}} 
\newcommand{\expmatch}{M}
\newcommand{\expmatchest}{\hat{M}}

\newcommand{\E}{\mathbb{E}}         

\newcommand{\algnamens}{MatchRank}
\newcommand{\algname}{\algnamens\ }  

\newtheorem{example}{Example}

\newcommand{\eq}[1]{Equation~\protect(\ref{#1})}

\newcommand{\alg}[1]{Algorithm~\protect\ref{#1}}
\newcommand{\fig}[1]{Figure~\protect\ref{#1}}

\newcommand{\theo}[1]{Theorem~\protect\ref{#1}}

\DeclareMathOperator*{\argmax}{argmax}
\DeclareMathOperator*{\argmin}{argmin}

{\makeatletter
 \gdef\xxxmark{%
   \expandafter\ifx\csname @mpargs\endcsname\relax 
     \expandafter\ifx\csname @captype\endcsname\relax 
       \marginpar{xxx}
     \else
       xxx 
     \fi
   \else
     xxx 
   \fi}
 \gdef\xxx{\@ifnextchar[\xxx@lab\xxx@nolab}
 \long\gdef\xxx@lab[#1]#2{{\bf [\xxxmark #2 ---{\sc #1}]}}
 \long\gdef\xxx@nolab#1{{\bf [\xxxmark #1]}}
}

\begin{document}


\title{Ranking with Slot Constraints}
\author{Wentao Guo\footnote{Equal contribution}, Andrew Wang\protect\footnotemark[1], Bradon Thymes,Thorsten Joachims}
\affil{\texttt{\{wg247, azw7, bmt63\}@cornell.edu, tj@cs.cornell.edu} \\
Cornell University}
\date{}

\setlist{nosep}
\newtheorem{definition}{Definition}
\newtheorem{remark}{Remark}
\newtheorem{lemma}{Lemma}
\newtheorem{observation}{Observation}
\newtheorem{theorem}{Theorem}

\DeclareDocumentCommand{\mathdef}{mO{0}m}{%
  \expandafter\let\csname old\string#1\endcsname=#1
  \expandafter\newcommand\csname new\string#1\endcsname[#2]{#3}
  \DeclareRobustCommand#1{%
    \ifmmode
      \expandafter\let\expandafter\next\csname new\string#1\endcsname
    \else
      \expandafter\let\expandafter\next\csname old\string#1\endcsname
    \fi
    \next
  }%
}

\maketitle

\begin{abstract}
       We introduce the problem of ranking with slot constraints, which can be used to model a wide range of application problems -- from college admission with limited slots for different majors, to composing a stratified cohort of eligible participants in a medical trial. We show that the conventional Probability Ranking Principle (PRP) can be highly sub-optimal for slot-constrained ranking problems, and we devise a new ranking algorithm, called \algnamens. The goal of \algname is to produce rankings that maximize the number of filled slots if candidates are evaluated by a human decision maker in the order of the ranking. In this way, \algname generalizes the PRP, and it subsumes the PRP as a special case when there are no slot constraints. Our theoretical analysis shows that \algname has a strong approximation guarantee without any independence assumptions between slots or candidates. Furthermore, we show how \algname can be implemented efficiently. Beyond the theoretical guarantees, empirical evaluations show that \algname can provide substantial improvements over a range of synthetic and real-world tasks.
\end{abstract}

\section{Introduction}


Rankings have become a ubiquitous interface whenever there is the need to focus attention among an otherwise impractically or intractably large number of options. Beyond their conception as an interface for query-based retrieval \cite[e.g.,][]{Salton/71a}, rankings are now widely used in related tasks like recommendation and advertising. However, a substantial number of new ranking applications come with additional constraints, and we identify the notion of {\em slot constraints} as a frequent requirement. With slot constraints we refer to capacity constraints for different types of relevant candidates. For example, there is only a certain number of slots for each major in a college admissions task; or the cohort of a medical trial may need to fulfill constraints on gender and race to be representative \cite{baquet2006recruitment}; and in a multi-stage retrieval pipeline we may have assortment constraints \cite{wang_improving_2022}. In these applications, the goal is to fill all slots with relevant candidates, and each candidate can have a different probability of relevance for each slot.

For rankings without constraints, the Probability Ranking Principle (PRP) \cite{robertson1977probability} has long been understood to provide the ranking that maximizes the number of relevant candidates that are found in the top-$k$ of the ranking, for any $k$. However, the PRP does not apply to ranking problems with {\em slot constraints}. We show that for ranking applications with slot constraints, the PRP can be highly sub-optimal, disparately spending human evaluation effort on candidates for which there are no open slots while ignoring other candidates that are relevant for unfilled slots. Furthermore, ranking with slot constraints is different from both intrinsically and extrinsically diversified ranking \cite{Radlinski/etal/09a}, since it allows us to put explicit constraints on the set of {\em relevant} results. Conventional diversification methods either ignore relevance (e.g., demographic parity \cite{linkedin/19a}), or do not allow the specification of constraints (e.g., \cite{carbonell1998use,chapelle2011intent,zhai2015beyond}).


To remedy this shortcoming, we formalize and address the problem of {\em ranking with slot constraints} in this paper. In our formulation, a human decision maker can define an arbitrary set of slots (e.g., admission slots for each major) that need to be filled with relevant candidates (i.e., qualified students). The ranker then supports the human decision maker in allocating the evaluation effort while leaving the relevance decisions to the human. Specifically, given a relevance model that estimates the relevance probability of each candidate for each slot, the goal is to compute a ranking under which the expected number of filled slots is maximized. This model is very general, as candidates can be qualified for any number of slots, and we can use any probabilistic model with arbitrary dependencies between the slot relevances of all candidates.  

Under this model, we derive a general ranking algorithm, called \algnamens, that merely requires efficient sampling from the relevance model. We theoretically analyze \algname and show that it provides a strong approximation guarantee. In particular, we prove that any top-$k$ of the ranking computed by \algname guarantees an expected number of filled slots that is asymptotically at most $(1-1/e)$ away from the optimal set with high probability. Furthermore, we show how \algname can be implemented efficiently to handle ranking problems of substantial size. Finally, we provide an extensive empirical evaluation of \algnamens, showing that it can outperform heuristic baselines by a substantial margin, and perform accurately on a real-world college-admission problem.

\section{Related Work}

In the following we detail how our new setting of ranking with slot constraints differs from existing research areas.








\textbf{Search result diversification} is a widely studied problem in IR in which one aims to cover multiple intents or aspects of an ambiguous or composite query. Specifically, in extrinsic diversification \cite{Radlinski/etal/09a} the goal it to cover all intents of a query to ensure that the user finds at least one relevant result despite the uncertainty about the query intent. This typically leads to coverage-style objectives \cite{Yue/Joachims/08a,carbonell1998use}, not matching problems like in slot-constrained ranking. For intrinsic diversification \cite{Radlinski/etal/09a}, the goal to to put together a portfolio of items, but none of the existing methods provides the same flexibility in specifying complex systems of slots with arbitrary dependencies. Another difference of our setting to existing diversification approaches is that we do not need to model similarity between items, either explicitly through predefined aspects \cite{santos2010exploiting} or implicitly through similarity metrics \cite{su2021modeling}. However, one could use our slot-constrained ranking framework as a method for diversification, especially in high-stakes selection problems where practitioner require both full transparency and control over the composition of the set of selected items, and the ability to specify complex real-world matching constraints.

Beyond the standard diversity settings, \citet{agarwal2015constrained} consider how to maximize user engagement while satisfying a minimum impression requirement per search result. However, their target objective is not a matching problem, since serving an item to one user does not exclude the use of such item with another user. Also related but different from our setting is the approach of \citet{dang2012diversity}. Their "seats" allocation approach is relevant to our slot constraints, except that we directly treat the slots as a constraint instead of a normalization factor. Therefore, we use the maximum bipartite matching algorithm \cite{hopcroft1973n} for the ranking objective. 



\textbf{Fairness in ranking} is also frequently implemented by adding constraints to the ranking, since ranking by predicted merit can lead to both poor representation \cite{arcidiacono2015affirmative} and suboptimal performance \cite{Ramist1994StudentGD} of admitted cohorts. In many cases, these fairness constraints ensure a certain amount of representation from different groups in various positions of the ranking (e.g., demographic parity) \cite{linkedin/19a,kleinberg_selection_2018}. This is critically different from our slot constraints, since slot constraints act on the {\em relevant} items, not all items. Note that under differential estimation accuracy between groups, merely ensuring representational fairness can still be unfair to the relevant items \cite{garg_dropping_2021,Singh/etal/21a}.


\textbf{Matching problems} have a wide range of applications in job markets, dating, and resource allocation in online clouds \cite{dickerson2019balancing,thekinen2017resource,khuller1994line}. The typical setting in \textit{stochastic matching} is that each edge in a graph is realized independently with (predicted) probability $p$ \cite{assadi2019stochastic,adamczyk2020improved}, which is analogous to the college admission scenario where we only have a calibrated regression model to know the predicted probability of a candidate $\cand$ being relevant to a slot $\slot$. \citet{dickerson2019balancing} consider online bipartite matching to improve the diversity and relevance of search results by maximizing a multilinear objective over the set of matched edges. \citet{ahmed2017diverse} propose a quadratic programming based objective for the diversity of a matching and propose a scalable greedy algorithm to trade off efficiency and diversity. Instead of evaluating an objective on top of matching on a sampled graph, we use the size of the bipartite matching on a sampled graph to derive a ranking of candidates that maximizes the size of the bipartite matching in the true relevance matrix during the human evaluation phase. We are the first to formulate this ranking problem to maximize the size of the bipartite matching, and this is a core contribution of this paper.

\section{Problem Setting}

Consider that we have $\CandidatesNum$ candidates $\Candidates=\{\cand_1, ... \cand_\CandidatesNum\}$, and we have $\SlotsNum<\CandidatesNum$ slots $\Slots=\{\slot_1, ...,\slot_\SlotsNum\}$ that we need to fill with relevant candidates. Each candidate can be relevant to any number of slots, or no slots at all. We denote whether candidate $\cand_i$ is relevant to slot $\slot_j$ via the matrix entry $\Rel_{ij} \in \{0,1\}$. We use the generic concept of "relevance" to indicate whether a candidate matches a slot. This allows us to model a broad range of selection problems as follows: 
\begin{description}
    \item[Hiring:] A company has a number of openings for different roles, with a specific number of interview slots budgeted for each role. Applicants may be qualified for some subset of the roles. An applicant is relevant for an interview slot if qualified and interested in the particular opening.
    \item[College Admission:] The slots correspond to seats in various majors (100 slots for CS, 50 slots for Math, etc.), and in each major a certain number of slots is reserved for low socioeconomic status students. A student is relevant for any slot in a major if both qualified for and interested in that major.
    \item[Medical Trial:] Researchers need to find qualifying participants for a medical trial among millions of electronic health records. The trial is designed with a certain number of slots by gender, race and disease severity. Patients are relevant for a slot if they match demographic requirements and disease severity determined by further testing. 
\end{description}
In all of these application scenarios our goal is to fill all slots with relevant candidates. Throughout this paper we assume that relevance is binary, but we conjecture that many of our results can be extended to non-binary relevance values.

If the relevance matrix $\Rel$ was fully known, the problem of finding relevant candidates to fill all slots would be solved by the maximum bipartite matching algorithm \cite{hopcroft1973n}. In practice, however, we only have uncertain information $P(\Rel)$ about the true relevances. $P(\Rel)$ will typically be a probabilistic relevance prediction model learned from data, and hence we merely assume access to $P(\Rel)$ instead of $\Rel$. Furthermore, accurately revealing the true relevance vector $\Rel_{i} \in \{0,1\}^\SlotsNum$ of any particular candidate comes at substantial cost. In the admissions and hiring examples, assessing relevance requires detailed human review of the application, and in the medical example it requires additional medical testing. We would thus like to avoid evaluating candidates that do not contribute to filling more slots, either because these candidates are not relevant or because we already have identified other relevant candidates. 

To achieve this goal, we would like to compute a ranking $\ranking$ of candidates so that evaluating the candidates $\ranki{1}, ... , \ranki{\CandidatesNum}$ from top to bottom maximizes the number of filled slots given the information contained in $P(\Rel)$. Without slot constraints (i.e., for any candidate $\cand_i$: $\forall j,k: \Rel_{ij} = \Rel_{ik}$) the ranking problem has a well-known solution that follows from the Probability Ranking Principle (PRP) \cite{robertson1977probability}: simply ranking the candidates by their probability of relevance is optimal under most sensible metrics. However, this PRP ranking can be highly sub-optimal under general slot constraints, as the following toy example shows. 

\begin{example}[Suboptimality of PRP for Ranking with Slot Constraints] Consider a problem with $\CandidatesNum=1000$ candidates and $\SlotsNum=10$ slots. Candidates  $\cand_1, ..., \cand_{500}$ have a probability of relevance of 0.5 for slots $\slot_1, ... ,\slot_5$, and 0 for the other slots. Analogously, candidates $\cand_{501}, ..., \cand_{1000}$ have a probability of relevance of 0.4 for slots $\slot_6, ... ,\slot_{10}$, and 0 for the other slots. Any heuristic based on sorting candidates by a score computed from their probability of relevance would either produce a ranking equivalent to $\cand_1,...,\cand_{500},\cand_{501},...,\cand_{1000}$ or equivalent to $\cand_{500},...,\cand_{1000},\cand_{1},...,\cand_{500}$. However, either ranking would be highly suboptimal, since one type of slots would not be filled until after reviewing at least 500 candidates.
\end{example}

Note that this high degree of suboptimality already surfaces in this particularly simplistic example, where there are only two types of slots and candidates are relevant to at most one slot. In the more general case, where we can have complex systems of slots where each candidate can have dependent probabilities of being relevant to multiple slots, it is not even clear how to heuristically apply the conventional PRP. This motivates the need for a new algorithm that goes beyond sorting candidates by some heuristic function of relevance, but that explicitly takes into account the slot constraints.
In the following we develop the \algname algorithm that does not have the inefficiencies of the PRP ranking and that provides provable guarantees on the quality of the ranking for arbitrary slot constraints and relevance models. To start the derivation, the following begins with a formal definition of the ranking objective.

\subsection{Ranking Objective}

We formalize the problem of ranking under slot constraints in two steps. We first define the problem of finding a candidate set $\candset_\candsetsize$ of a given size $\candsetsize$ that is optimal. In the second step we show how to construct a nested sequence 
\begin{equation}
    \candset_0 \subset \candset_1 \subset \candset_2 \subset ... \label{eq:nestedsets}
\end{equation}
of such candidate sets that naturally forms a ranking $\ranking$. In particular, any two consecutive candidate sets $\candset_{k}$ and $\candset_{k+1}$ differ by only one candidate $\candset_{k+1} \setminus \candset_{k} = \cand_\ranki{k+1}$, which corresponds to the element $\ranki{k+1}$ of ranking $\ranking$. Note that $\candset_0$ is always the empty set.

\begin{figure}[t] 
    \centering
    \includegraphics[width=0.65\columnwidth]{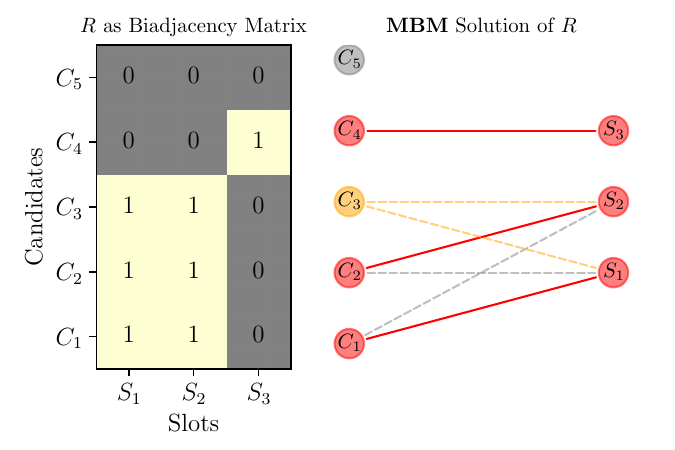}
    \vspace*{-0.3cm}
    \caption{Example showing how MBM computes an optimal assignment of candidates to slots for a known relevance matrix $\Rel$. In this figure, \textcolor{gray}{$C_5$} is not relevant for any slots. \textcolor{red}{$C_1$}, \textcolor{red}{$C_2$} and \textcolor{red}{$C_4$} are relevant and can be matched with available slots. \textcolor{orange}{$C_3$} is relevant for $S_1$ and $S_2$, but both are already occupied.} 
    \label{fig:bipartitematching}
\end{figure} 

To evaluate a given candidate set $\candset \subseteq \Candidates$, we use the size of the maximal matching between relevant candidates in $\candset$ and slots $\Slots$. This is illustrated in \fig{fig:bipartitematching}, where the candidates in $\candset$ and the slots in $\Slots$ form a bipartite graph (right panel). The corresponding submatrix of the relevance matrix $\Rel$ (left panel) defines the biadjacency matrix of the graph, where an edge exists whenever a candidate $\cand_i$ is relevant for slot $\slot_j$. The right panel of \fig{fig:bipartitematching} shows the (not necessarily unique) maximum bipartite matching $\{ \cand_1 \rightarrow \slot_1, \cand_2 \rightarrow \slot_2, \cand_4 \rightarrow \slot_3 \}$, which corresponds to the largest number of slots that can be filled with relevant candidates from $\candset$. We denote this maximum bipartite matching size with $\mbm{\candset}{\Slots}{\Rel}$.


While $\mbm{\candset}{\Slots}{\Rel}$ gives us the optimal solution for a known relevance matrix $\Rel$, we need to evaluate candidates sets $\candset$ under uncertainty about what the correct relevance matrix is. A natural metric for evaluating a candidate set $\candset$ under $P(\Rel)$ is to measure the expected size of the matching, which corresponds to the expected number of slots that can be filled with candidates from $\candset$.
\begin{eqnarray}
    \expmatch(\candset) &=& \E_{\Rel \sim P(\Rel)} \bigl[ \mbm{\candset}{\Slots}{\Rel} \bigr] \\
                        &=& \sum_{\Rel} \mbm{\candset}{\Slots}{\Rel} P(\Rel) \label{eq:expmatch}
\end{eqnarray}
When evaluating a ranking $\ranking$, we will apply this metric $\expmatch(\candset)$ to each top-$k$ prefix $\candset_k$ of the ranking $\ranking$.

\subsection{\algname Algorithm} \label{sec:algorithm}

Given the metric $\expmatch(\candset)$ from Equation~\eqref{eq:expmatch}, our goal is to find a ranking $\ranking$ of the candidates in $\Candidates$ so that $\expmatch(\candset_k)$ for any top-$k$ prefix is as large as possible under $P(\Rel)$. We split this goal into three steps. First, we show how to estimate $\expmatch(\candset)$ for any candidate set $\candset$ and for any $P(\Rel)$ that permits sampling. Second, we show how to construct a single candidate set $\candset_k$ of size $\candsetsize$ that has a large value for $\expmatch(\candset_k)$. Third, we show that our construction of the $\candset_k$ in the previous step naturally produces a ranking.

\paragraph{Estimating $\expmatch(\candset)$.} We do not make any structural assumptions on the distribution $P(\Rel)$, and $P(\Rel)$ can contain arbitrary dependencies between the entries. The following is a general method for evaluating a given candidate set $\candset$, where we merely require that we can sample relevance matrices $\Rel$ from $P(\Rel)$. With such samples, we can compute Monte-Carlo estimates of $\expmatch(\candset)$ as follows.

Let $\RelSamples= [\Rel_1,...,\Rel_\n]$ be $\n$ samples of relevance matrices drawn i.i.d. from $P(\Rel)$. The Monte-Carlo estimate of $\expmatch(\candset)$ is
\begin{equation}
    \expmatchest(\candset) = \frac{1}{\n}\sum_{i=1}^{\n} \mbm{\candset}{\Slots}{\Rel_i}. \label{eq:expmatchest}
\end{equation}
By the weak law of large numbers, $\expmatchest(\candset)$ converges to $\expmatch(\candset)$ for large $\n$. We will later characterize how the number of samples $\n$ affects the algorithm.

\paragraph{Constructing the ranking.} Now that we know how to estimate the quality $\expmatchest(\candset)$ of any particular candidate set $\candset$, we can think about finding a candidate set that maximizes $\expmatchest(\candset)$. However, naively enumerating all subsets $\candset \subset \Candidates$ of size $|\candset| = \candsetsize$ and evaluating $\expmatchest(\candset)$ would not be efficient. Furthermore, it would not be clear how to make sure that the candidate sets are nested and from a ranking.

We avoid this combinatorial enumeration and instead construct $\expmatchest(\candset)$ using the following greedy algorithm, which we will prove to enjoy strong approximation guarantees. Since this algorithm adds one candidate in each iteration, it naturally constructs a ranking and we show that the approximation guarantees hold for any top-$k$ of the ranking.

\begin{algorithm}[ht]
    \caption{MatchRank}\label{alg:greedy}
    \begin{algorithmic}
    \Require \, candidates $\Candidates$; slots $\Slots$; sampled relevances $\RelSamples= [\Rel_1,...,\Rel_\n]$; 
        \State $\candset_0 \gets \emptyset$; $A \gets \Candidates$; $\ranking = []$; $k \gets 1$
        \While{$A \neq \emptyset$} 
            \State $\cand_{\text{best}} \gets \text{argmax}_{\cand \in A} \frac{1}{\n} \!\!\sum_{j=1}^{\n} \!\mbm{\!\candset_{k-1} \!\!\cup\! \{\cand\}}{\Slots}{\Rel_j\!}$           
            \State $\ranki{k} =\cand_\text{best}$ 
            \State $\candset_{k} \gets \candset_{k-1} \cup \{\cand_\text{best}\}$; $A \gets A - \{\cand_\text{best}\}$; $k \gets k+1$       
        \EndWhile
    \Ensure \, ranking $\ranking$
    \end{algorithmic}
\end{algorithm}

In each iteration $k$, the \algname algorithm finds the candidate $\cand_\text{best}$ that most improves $\expmatchest(\candset_{k-1} \cup \{\cand_\text{best}\})$ for the current candidate set $\candset_{k-1}$. It then places $\cand_\text{best}$ into position $k$ of the ranking. Furthermore, it adds $\cand_\text{best}$ to the current top-$k$ set $\candset_{k}$, and it removes $\cand_\text{best}$ from the set of remaining candidates $A$. These iterations continue until all candidates have been added to the ranking. If only a top-$k$ ranking was desired, one could also stop early. Note that Algorithm \ref{alg:greedy} is optimized for clarity, but the supplementary material presents several efficiency improvements.


\subsection{Theoretical Analysis} \label{sec:theory}

We now analyze theoretically how effective \algname is at constructing a ranking that optimizes the objective $\expmatch(\candset)$ given in \eq{eq:expmatch}. We start by stating our main result, which we will then prove subsequently. The main result states that for any $k$, the top-$k$ candidate set $\candset_k$ constructed by \algname enjoys an approximation guarantee compared to the optimal candidate set $$\candset_k^* = \argmax_{\candset \subseteq \Candidates \wedge |\candset|=k} \expmatch(\candset).$$ Note that these optimal $\candset_k^*$ may not be nested and may not form a ranking, unlike the $\candset_k$ constructed by \algnamens.

\begin{theorem}\label{thm:exp}
     The ranking $\ranking$ produced by \algname when given $\SlotsNum$ slots and $\n$ Monte-Carlo samples $\RelSamples= [\Rel_1,...,\Rel_\n]$ from $P(\Rel)$ enjoys the following approximation guarantee for each top-$k$ set $\candset_k$ in $\ranking$: with probability $1-\delta$ (where $0<\delta<1/2$),
    \[ M(\candset_k) \geq \left(1-\frac{1}{e}\right) M(\candset_k^*) - 2 s\sqrt{\frac{O(k\ln k) + \ln(2/\delta)}{2n}}, \]
    where $\candset_k^* = \argmax_{\candset \subseteq \Candidates \wedge |\candset|=k} \expmatch(\candset)$ is the optimal set.
\end{theorem}

The proof of \theo{thm:exp} is given in the supplementary material. Its main steps are to first show that $\expmatchest(\candset)$ is monotone submodular, which implies that the greedy nature of \algname provides a $(1-1/e)$ approximation guarantee for $\expmatchest(\candset)$. We then show that optimizing $\expmatchest(\candset)$ provides a solution that is close to optimizing $\expmatch(\candset)$ directly.

\usetikzlibrary{automata,positioning}
\tikzstyle{startstop} = [rectangle, rounded corners, text centered, draw=black, fill=red!30]
\tikzstyle{process} = [rectangle, text centered, draw=black, fill=orange!30]
\tikzstyle{slots} = [rectangle, rounded corners, text centered, draw=black, fill=blue!30]
\tikzstyle{decision} = [rectangle, text centered, draw=black, fill=green!30]
\tikzstyle{ranker} = [diamond, text centered, draw=black, fill=green!30]
\tikzstyle{R} = [rectangle,  rounded corners, text centered, draw=black, fill=yellow!30]
\tikzstyle{arrow} = [thick,->,>=stealth]
\tikzstyle{dasharrow} = [thick,->,-]

\newcommand{\FixedLengthArrow}{2,0}

\begin{figure*}[t]
    \begin{tikzpicture}[>=stealth, on grid, auto]
    \node (model) [process] { Regression models};
    \node (data) [startstop, below of=model, yshift=2cm] {Candidates features};
    \node (pr) [process, right of=model, xshift=1.3cm] {$\PR$};
    \node (samples) [process, right of=pr, xshift=1.8cm, text width=3.0cm, line width=0.1pt] {Sampled relevance matrices $\RelSamples=\{\Rel_1,...,\Rel_n\}$};
    \node (ranker) [decision, right of=samples, xshift=2.3cm, text width=2cm] {Ranking algorithm };
    \node (ranking) [decision, right of=ranker, xshift=1.7cm, text width=2cm] {Ranking shortlist $\ranking$};
    \node (slots) [slots, below of=ranker, yshift=2cm] {Allocated slots};
    \node (human) [startstop, right of=ranking, xshift=2.4cm, text width=3.5cm] {Find \# filled slots by relevant candidates from $\ranking$};
    \node (R) [R, below of=human, yshift=2cm] {Ground truth relevances $\Rel$};
    \draw [arrow] (data) -- (model);
    \draw [arrow] (model) -- (pr);
    \draw [arrow] (pr) -- (samples);
    \draw [arrow] (samples) -- (ranker);
    \draw[bend left=35,->]  (pr) to node [auto] {} (ranker);
    \draw [arrow] (slots) -- (ranker);
    \draw [arrow] (ranker) -- (ranking);
    \draw [arrow] (ranking) -- (human);
    \draw [arrow] (R) -- (human);
    \end{tikzpicture}
    \caption{Application and evaluation pipeline used in the ranking experiments. \label{fig:pipeline} } 
\end{figure*}
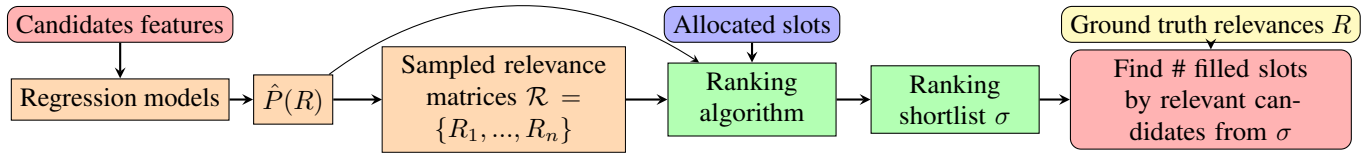

\begin{table*}[!ht] 
 \centering 
 \setlength{\tabcolsep}{3.5pt}

\small

 \begin{tabular}{l | c | cc | cc | cc | cc r}
    \toprule
    
    & \textbf{Default} &
    \multicolumn{2}{c|}{\textbf{\# Slots Per Group}} &  
    \multicolumn{2}{c|}{\textbf{\# Group Memberships}} &
    \multicolumn{2}{c|}{\textbf{\# Samples}} &
    \multicolumn{2}{c}{\textbf{$P(R)$}} 
    \\ \hline
    
    & 
    & 30 & 70  
    & 1 & 3 
    & 100 & 1000
    & S & L
    \\ \hline
    
    \midrule
    
    \textbf{\textit{MatchRank}} 
    & \textbf{1.27 $\pm$ 0.06} 
    & \textbf{1.26 $\pm$ 0.07} & \textbf{1.29 $\pm$ 0.05}  
    & \textbf{2.05 $\pm$ 0.20} & \textbf{1.12 $\pm$ 0.03} 
    & \textbf{1.32 $\pm$ 0.10} & \textbf{1.25 $\pm$ 0.03} 
    & \textbf{1.52 $\pm$ 0.13} & \textbf{1.14 $\pm$ 0.03} 
    \\

    \textbf{AND} 
    & 5.01 $\pm$ 0.29 
    & 6.06 $\pm$ 0.64 & 4.29 $\pm$ 0.20
    & 11.26 $\pm$ 0.31& 2.73 $\pm$ 0.24
    & 5.01 $\pm$ 0.29 & 5.00 $\pm$ 0.28
    & 5.84 $\pm$ 0.39 & 4.45 $\pm$ 0.25
    \\
    
    \textbf{OR} 
    & 4.20 $\pm$ 0.33 
    & 4.56 $\pm$ 0.53 & 3.85 $\pm$ 0.22
    & 11.26 $\pm$ 0.31& 2.29 $\pm$ 0.21
    & 4.21 $\pm$ 0.34 & 4.20 $\pm$ 0.33
    & 5.31 $\pm$ 0.41 & 3.29 $\pm$ 0.29
    \\

    \textbf{TR} 
    & 4.41 $\pm$ 0.31 
    & 5.48 $\pm$ 0.39 & 4.00 $\pm$ 0.23
    & 10.73 $\pm$ 0.44& 2.57 $\pm$ 0.17
    & 4.14 $\pm$ 0.31 & 4.73 $\pm$ 0.31
    & 5.46 $\pm$ 0.41 & 3.90 $\pm$ 0.26
    \\
    
    \textbf{NTR} 
    & 1.35 $\pm$ 0.07 
    & 1.45 $\pm$ 0.12 & 1.30 $\pm$ 0.04
    & 3.69 $\pm$ 0.14 & \textbf{1.12 $\pm$ 0.02}
    & 1.45 $\pm$ 0.07 & 1.39 $\pm$ 0.09
    & 1.66 $\pm$ 0.13 & 1.20 $\pm$ 0.05  
    \\

    \textbf{Random} 
    & 1.69 $\pm$ 0.11 
    & 1.78 $\pm$ 0.18 & 1.68 $\pm$ 0.02
    & 3.70 $\pm$ 0.40 & 1.23 $\pm$ 0.04
    & 1.69 $\pm$ 0.11 & 1.69 $\pm$ 0.11
    & 2.51 $\pm$ 0.25 & 1.35 $\pm$ 0.07
    \\
    
    \bottomrule
    \end{tabular}

    \caption{Synthetic Datasets: Performance of \algname in comparison to the heuristic baselines, reporting mean and standard deviation (for standard error divide by $\sqrt{1000}$) of $k_{\min} / | \Slots|$ over 1000 random draws of the true relevance matrix $\Rel$ from $P(\Rel)$.}
    \label{tab:synthetic}
    
\end{table*}

Note that the $\sqrt{.}$ term approaches zero as we increase the number $\n$ of sampled relevance matrices. This means that for large Monte-Carlo samples, the approximation factor approaches $1-1/e$, where $e$ is Euler's number. 

    
    
    
    

The monotone submodularity of $\expmatchest(\candset)$ opens a large arsenal of submodular optimization methods for constructing candidate sets $\candset_k$ with provable approximation guarantees. We opted for the greedy algorithm, since it naturally constructs a ranking.

\section{Empirical Evaluation}

We now evaluate the \algname algorithm on three types of data. The first is fully synthetic data, where we can control all aspects of the ranking problems to understand the conditions under which \algname improves over baseline heuristics. Second, we evaluate \algname on a number of benchmark datasets. And finally, we verify the applicability of \algname on a real-world college-admissions problem. 

\paragraph{Evaluation Process and Metric} For each problem, our evaluation follows the process depicted in Figure~\ref{fig:pipeline}. We first use a training set to learn a model --- usually a calibrated regression model -- which we can use to infer the relevance probabilities $\PR$ for the candidates in the test set. We apply \algname and other baseline rankers to rank this test set, which only requires sampling from $\PR$. 

To evaluate any ranking $\ranking$, we use the following process and metric. For each top-$k$ prefix $\ranking_k$ of $\ranking$, we reveal the ground-truth relevance labels $\Rel$ of these $k$ candidates and compute how many slots can be filled when optimally matching these $k$ candidates to the slots. This is precisely the size of the matching $MBM(\ranking_k, \Slots | \Rel)$. Our final evaluation metric for $\sigma$ is the smallest $k$ for which the prefix $\ranking_k$ fills all $|\Slots|$ slots with relevant candidates.
\begin{equation}
    k_{\min}=\argmin_{k \in [c]} \left\{ MBM(\ranking_k, \Slots | \Rel) = | \Slots | \right\}
\end{equation}
This means that $k_{\min}$ is the number of candidates in $\ranking$ that need to be reviewed before all slots are filled.

Since $k_{\min}$ scales with the total number of slots, we report the normalized $k_{\min} / |\Slots|$, so that the best possible score is 1.

\paragraph{Baseline Rankers}\label{para:benchmarks}

Since no prior methods exists for our problem, we created the following baselines. These methods compute a score for each candidate, and then rank by this score. The baselines rankers differ by how they aggregate the estimated marginal relevance probabilities $\hat{P}(R_{\cand, \slot})$ (estimated from the Monte-Carlo samples) for each candidate $\cand$ across all slots $\slot$. The first heuristic is motivated by the soft \textbf{AND} rule $\prod_{S \in \Slots} \hat{P}(R_{\cand, S})$. The second uses the soft \textbf{OR} rule $1 - \prod_{S \in \Slots} (1 - \hat{P}(R_{\cand, S})) $. Both the \textbf{AND} and the \textbf{OR} rule skip probabilities that are zero when computing the product. The third heuristic, called Total Relevance (\textbf{TR}) merely sums the relevance probabilities $\sum_{S \in \Slots} \hat{P}(R_{\cand, S})$ across all slots. The final heuristic is a normalized version of the total relevances, called \textbf{NTR}, that normalizes with the competition for each slot $\sum_{S \in \Slots} \frac{\hat{P}(R_{\cand, S})}{\sum_{\cand' \in \Candidates} \hat{P}(R_{\cand', S})}$.

\subsection{Synthetic Datasets}

We first focus on a synthetic dataset where we can control the structure of $P(R)$, so that we can investigate all problem dimensions that affect the \algname algorithm. Furthermore, we can directly use $P(R)$ instead of $\PR$, which avoids confounding the algorithm's behavior with potential inaccuracies in a learned $\PR$.


\paragraph{Experiment Setup} \label{para:synthetic-settings}


To construct synthetic $P(R)$, we define $g$ groups (default $g=10$) and each group has $s$ slots (default $s=50$). We then create 10,000 candidates, where each candidate is randomly assigned to $a$ groups (default $a=2$). If candidate $\cand$ is a member of group $j \in \{1,...,g\}$, we first sample $p$ from the Gaussian distribution $\mathcal{N}(p_\text{base} + 0.03 * j, 0.1)$ with default $p_\text{base}$ as 0.3, and clip $p$ to the range of (0.0001, 0.9999). We then sample from a Bernoulli distribution with probability $p$, and the success outcome means $\cand$ is relevant for all slots associated with group $j$. If candidate $\cand$ is not member of group $j$,  then $\cand$ is not relevant for any slots associated with group $j$, or formally $P(R_{\cand, S}) = 0$ if slot $S$ belongs to group $j$. If not mentioned otherwise, parameters are at their default value.

We draw 200 Monte-Carlo i.i.d. samples $\RelSamples=[R_1,...,R_{200}]$ as input to the ranking algorithms. For evaluation, we draw a ground-truth relevance matrix $\Rel$ and compute $k_{\min} / |\Slots|$, repeat this evaluation 1000 times, and report the mean and standard deviation.

\begin{table*}[!ht]
    \setlength{\tabcolsep}{3.6pt}
    \small
    \begin{tabular}{l | ccc | ccc | ccc | ccc | ccc | ccc r}

    \toprule
    \textbf{Datasets} 
    & \multicolumn{3}{c|}{\textbf{Medical}} & 
    \multicolumn{3}{c|}{\textbf{Bibtex}} & 
    \multicolumn{3}{c|}{\textbf{Delicious}} & 
    \multicolumn{3}{c|}{\textbf{TMC2007}} & 
    \multicolumn{3}{c|}{\textbf{Mediamill}} & 
    \multicolumn{3}{c}{\textbf{Bookmarks}} 

    \\ \hline
    \textbf{Slots Per Label} 
    & 5 & 10 & 15   
    & 10 & 20 & 30  
    & 10 & 30 & 50 
    & 30 & 50 & 70  
    & 10 & 30 & 50  
    & 10 & 30 & 50  

    \\ \hline
    \midrule
    
    \textbf{\textit{MatchRank}} 
    & \textbf{1.96} & \textbf{1.86} & \textbf{1.84}   
    & \textbf{3.17} & \textbf{2.27} & \textbf{2.19}  
    & \textbf{1.07} & \textbf{1.09} & \textbf{1.18}   
    & \textbf{1.17} & \textbf{1.32} & \textbf{1.26} 
    & \textbf{1.56} & \textbf{1.04} & \textbf{1.10} 
    & \textbf{5.03} & \textbf{4.09} & \textbf{3.53}
    \\

    \textbf{AND} 
    & 3.30 & 2.97 & 2.47   
    & 8.76 & 5.26 & 4.00 
    & 2.30 & 2.01 & 1.68  
    & 11.63 & 8.26 & 6.48
    & 6.15 & 2.17 & 2.08 
    & 66.01 & 23.05 & 24.83
    \\
    
    \textbf{OR} 
    & 3.42 & 3.25 & 2.59   
    & 6.71 & 4.51 & 3.61  
    & 2.74 & 2.21 & 1.99 
    & 3.40 & 3.49 & 3.02
    & 13.29 & 3.01 & 2.35 
    & 22.32 & 11.82 & 9.88
    \\

    \textbf{TR} 
    & 4.06 & 2.73 & 2.50  
    & 6.36 & 4.53 & 3.61
    & 3.15 & 2.21 & 2.03  
    & 8.61 & 6.13 & 5.09
    & 10.73 & 2.41 & 2.14 
    & 22.46 & 12.46 & 9.62
    \\
    
    \textbf{NTR} 
    & 4.42 & 2.60 & 1.99   
    & 6.06 & 3.40 & 2.41
    & 1.71 & 1.76 & 1.55  
    & 1.88 & 1.58 & 1.47
    & 2.83 & 1.62 & 1.54 
    & 20.81 & 10.08 & 6.99
    \\

    \textbf{Random} 
    & 4.14 & 3.70 & 3.53   
    & 8.18 & 6.36 & 6.37  
    & 1.66 & 1.93 & 2.06 
    & 4.31 & 3.72 & 3.39
    & 4.36 & 2.89 & 2.81  
    & 13.66 & 12.50 & 13.65
    \\
    
    \bottomrule

    \end{tabular}
    
    \caption{Real-World Benchmarks: Performance of \algname in comparison to heuristic baselines in terms of $k_{\min} / | \Slots|$.}
    \label{tab:multi-label-rank-size}
\end{table*}

\paragraph{Comparing \algname against the Baselines}

The first column of Table~\ref{tab:synthetic} shows the performance of \algname and the baselines for the default values of the synthetic data generator. \algname achieves a performance of 1.27, which means that on average only an additional 27\% of candidates need to be reviewed beyond the number of slots. The best heuristic is NTR, which averages 35\%. Most heuristics do worse than random, which requires 69\%. The reason is that the heuristics systematically miss candidates of some group, such that those slots cannot be filled.

\paragraph{Effects of the Number of Slots Per Group}

We now vary the number of slots per group from the default of 50 to 30 and 70. The results for 30 and 70 are in the third main column of Table~\ref{tab:synthetic}, while the results for 50 are in the default column. All other parameters are at their default values. \algname again show stable performance over all three settings and dominates most baselines. Only NTR comes close for larger numbers of slots.


\paragraph{Effects of Number of Group Memberships}

We now assign each candidate to be a member of 1, 2 (default) and 3 groups respectively. All other parameters are at their default. The fourth main column of Table~\ref{tab:synthetic} shows that the problem becomes easier for all methods, when each candidate can be in more groups. We see that the advantage of \algname is greater for harder problems.


\paragraph{Effect of Number of Monte-Carlo Samples}

We now investigate how important the number $n$ of Monte-Carlo samples $\RelSamples$ is, as we vary $n$ between 100, 200 (default), and 1000. The results can be found in the fifth main column of Table~\ref{tab:synthetic}. We find that \textit{MatchRank} already performs well for $n=100$, although having larger samples still improve the results. A larger $n$ may be even more important for $P(\Rel)$ with stronger dependencies between candidates and slots.

\paragraph{Effect of Overall Relevance Level}

We now vary the overall probability of relevance. In our model, this is controlled by the parameter $p_\text{base}$, and the higher $p_\text{base}$ the greater the overall probability of relevance. We vary $p_\text{base}$ between 0.2 (S), 0.3 (default), and 0.4 (L). The results can be found in the last major column of Table~\ref{tab:synthetic}. As $p_\text{base}$ increases, the problem of finding relevant candidates to fill the slots becomes easier. Both \algname and the baselines benefit from this, but \algname maintains a consistent advantage.

\paragraph{Additional results} in the supplementary material further support and elaborate on these findings.

\subsection{Real-World Benchmark Datasets} \label{sec:multilabel}


We now evaluate \algname on six benchmark datasets, where we now learn the relevance model $\PR$ from training data. These benchmark datasets are constructed from the multi-label datasets Medical \cite{pestian2007shared}, Bibtex \cite{katakis2008multilabel}, Delicious \cite{tsoumakas2008effective}, TMC2007 \cite{srivastava2005discovering}, Mediamill \cite{snoek2006challenge}, and Bookmarks \cite{katakis2008multilabel} from the Mulan data repository \cite{tsoumakas2009mining}. Our data source is the scikit-multilearn library \cite{2017arXiv170201460S}. 

\paragraph{Experiment Setup}

Each dataset comes with a train/test split. We consider each example in test set as a candidate, and each label as a group. We first select 10 labels that cover different numbers of relevant candidates, and then allocate a certain number of slots for each label. We train $\PR$ on the training set using a binary logistic regression for each label. To raise the noise in the relevance prediction to a more challenging and realistic level, we mask 20\% of the label in both the train and test set. We then draw 100 Monte-Carlo samples from $\PR$ as input to the ranking algorithm. All rankers are evaluated against the true masked relevance labels in the test set, meaning that a test example matches a slot if it contains the corresponding label.


\begin{figure}[t]
    \centering
    \includegraphics[width=0.65\columnwidth]{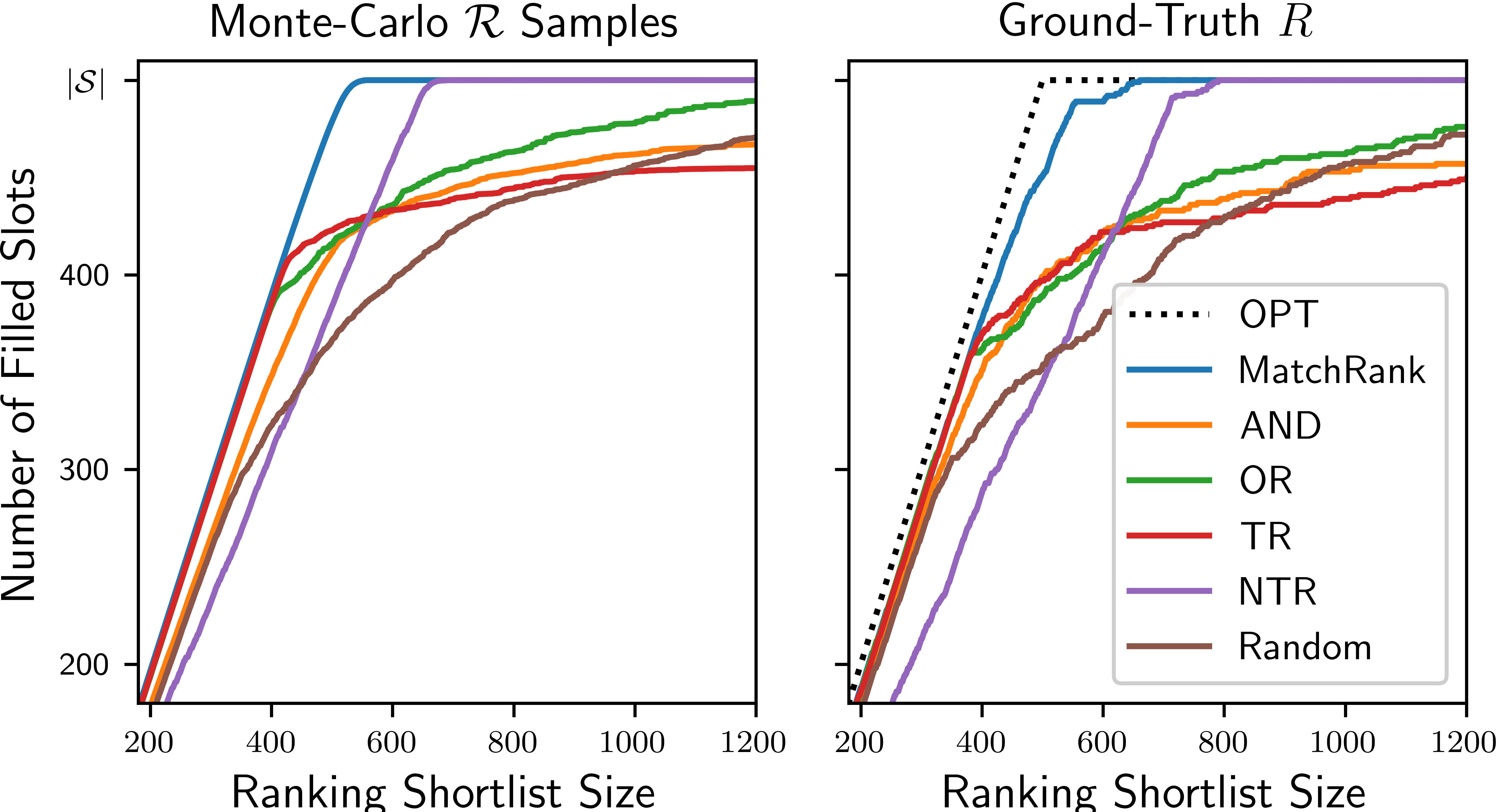}
    \vspace*{-0.3cm}
    \caption{TMC2007 Dataset: Number of filled slots on the Monte-Carlo samples (left) versus the expected number of filled slots w.r.t.\ the ground-truth relevances $\Rel$ (right). "OPT" is the optimal ranking if the ground-truth $\Rel$ was known, which achieves $MBM(\ranking_{|\Slots|}^{OPT}, \Slots | \Rel) = |\Slots|$. \label{fig:multi-label-matching-score}}
\end{figure}

\paragraph{Results}

The result are shown in Table~\ref{tab:multi-label-rank-size}. For all benchmark datasets and all numbers of slots per label, \algname delivers the best ranking performance in terms of $k_{\min}/|\Slots|$. The most competitive heuristic ranker is NTR, but the results show that its heuristic can fail on some datasets and provide substantially worse ranking performance than \algname (e.g. Bookmarks). These results demonstrate that \algname performs robustly over a wide range of datasets where the $\PR$ is learned from training data.

\paragraph{Analysis}

To provide additional insights into the behavior of \algname in comparison to the heuristic rankers, Figure~\ref{fig:multi-label-matching-score} shows their behavior on the TMC2007 dataset. The plot on the left shows the average number of filled slots over the Monte-Carlo samples as the shortlist size $k$ grows. With a total of 500 slots, \algname finds a close to optimal ranking while all heuristics require a substantially longer shortlist. This indicates that the heuristics optimize a fundamentally wrong objective. The plot on the right of Figure~\ref{fig:multi-label-matching-score} shows the number of filled slots when using the ground-truth labels for matching. This plot shows a gap between \algname and the optimal ranking one can compute from the ground-truth labels. Since \algname performs close to optimal on the Monte-Carlo samples, this gap can be attributed to the inherent inaccuracy and uncertainty of $\PR$ in predicting the ground-truth relevance. This suggests that \algname performs close to optimal in terms of its optimization performance, and that the remaining suboptimality is largely a result of an imperfect $\PR$.

\subsection{College Admission Dataset} \label{sec:admission}

To verify the effectiveness of \algname under real-world conditions, we consider an anonymized undergraduate admission dataset from a selective US university. The groups in this datasets are majors, and we posit that each major has only a fixed number of slots for admitting qualified students. We consider all majors that admitted at least 50 students, which leaves us with 13 majors and 19421 applicants in the test set. On the admission decisions from the prior year we train a boosted tree model with XGBoost \cite{xgboost} using a logistic loss objective and L2-regularization. This model is used to predict each applicant's probability of being admitted, and we clip the maximum $\hat{P}(R)$ by 0.3. Applicants can indicate interest in majors, and the probability of admission to a major that the applicant did not indicate interest in is set to zero. This provides us with $\PR$ for all test applicants.

We create 100 Monte-Carlo sampled relevance matrices from this $\PR$ as input to all ranking algorithms. To set the number of available slots for each member, we use $\min( \lfloor0.7 * |$relevant applicants for this major$| \rfloor, s_{\text{max}})$ to get interesting relationship between supply and demand, and we vary $s_\text{max}$ in the following experiments. We then run our ranking algorithms, and during evaluation we reveal the applicants ground-truth admissions decisions (i.e. the true relevance matrix $\Rel$) for each candidate. 
To evaluate, we again find the minimum shortlist size $k$ for each algorithm at which all slots are filled with relevant applicants. 

The results are provided in Table~\ref{tab:admission-rank-size}. This dataset is substantially more challenging for all methods, as the density of relevant candidates is small and we need to find a larger fraction of the relevant candidates to fill all slots. However, even in this challenging setting, we see that \algname is more effective than the heuristic baselines. This holds particularly when the number of slots per major is smaller, as the performance gap between \algname and other heuristic algorithms becomes larger. For larger $s_{\max}$ the majority of relevant candidates needs to be found, such that even a small degree of inaccuracy in $\PR$ can have large impact.

\begin{table}[t]
 \setlength{\tabcolsep}{9pt}
 \small
 \centering
     \begin{tabular}{l |cccc r}
    \toprule    
    \textbf{$s_\text{max}$} & 30 & 50 & 70 & 100

    \\ \hline
    \midrule
    \textbf{\textit{MatchRank}} 
    & \textbf{7.23} & \textbf{8.53} & \textbf{9.00} & \textbf{11.00} 
    \\
    
    \textbf{AND} & 19.63 & 11.91 & 15.77 & 11.98
    \\

    \textbf{OR} & 26.28 & 24.01 & 22.25 & 20.44
    \\

    \textbf{TR} & 29.98 & 22.40 & 18.58 & 15.82 
    \\
    
    \textbf{NTR} & 9.50 & 11.87 & 11.61 & 11.50 
    \\
    
    \textbf{Random} & 42.08 & 31.42 & 26.13 & 22.26 
    \\
    
    \bottomrule
    \end{tabular}
    
    \caption{College Admission: Performance of \algname in comparison to heuristics in terms of $k_{\min} / | \Slots|$.}
    \label{tab:admission-rank-size}
\end{table}

\section{Extensions and Future Work}

Instead of binary relevances as assumed in this paper, some applications may require real-valued relevances (e.g., star-ratings) where the decision maker aims to maximize the sum of relevances under slot constraints. In this setting, the optimal solution is given by the Maximum Weight Bipartite Matching $MWBM(X, \Slots | \Rel)$ for the weighted biadjacency matrix $\Rel$. If the MWBM objective is also monotone submodular, we could simply replace the MBM in \algname with the MWBM and provide a similar approximation guarantee for this weighted version of \algnamens.



Another extension is the use of high-dimensional matchings instead of the bipartite matching considered in this paper. This could model slot constraints over more than one category. For example, college admission may have slot constraints not only for majors but also for extracurricular teams (e.g., orchestra, athletics). Since slots for majors are orthogonal to the slots for extracurricular teams, this corresponds to a three-dimensional matching (3-DM) problem. The 3-DM problem is NP-hard and even APX-hard \cite{hazan2003complexity}, and the best approximation algorithm so far achieves an error bound of (4/3 + $\epsilon$) \cite{cygan2013improved}. In general, \citet{hazan2003complexity} showed that all $d$-DM ($d \geq 3$) problems cannot be approximated within a factor of $O( d /\ln d )$ unless P = NP. But even if this matching problem cannot be solved exactly any more, the approximate solutions could give rise to an approximate ranking algorithm analogous to \algnamens.

\section{Conclusion}
We introduce the problem of ranking under slot constraints, which allows practitioner to specify conditions that arise in a wide variety of applications. To solve this ranking problem, we develop the \algname algorithm and show that it provides a theoretical guarantee on its ranking performance. A key insight is that the ranking objective can be related to bipartite matching problems, and that it is monotone submodular. We also show how \algname can be implemented efficiently so that it can efficiently handle real-world ranking problems of substantial size. Beyond its theoretical guarantees, \algname shows superior ranking performance across extensive experiments compared to several heuristic baselines. This holds across a wide range of datasets and experiment conditions, and \algname shows robustness to sample size and misspecified relevance distributions. 
We conclude that the ability to model complex problems and provide accurate rankings across a wide range of domains, backed by theoretical guarantees, makes the slot constraint framework a promising paradigm for tackling complex real-world ranking problems.

\section{Acknowledgments}

This research was supported in part by NSF Awards IIS-1901168, IIS-2008139 and IIS-2312865. All content represents the opinion of the authors, which is not necessarily shared or endorsed by their respective employers and/or sponsors.

\bibliography{reference}
\bibliographystyle{plainnat}

\appendix

\newpage
\mbox{}

\section{Supplementary Material}

\subsection{Proof of Theorem~\ref{thm:exp}}


The first step in proving Theorem~\ref{thm:exp} is to prove that the maximum bipartite matching is monotone submodular.

\begin{lemma} \label{lem:submodmbm}
For any relevance matrix $\Rel$ and any set of slots $\Slots$, the size of the maximum bipartite matching is monotone in $\candset$
\begin{equation*}
    \forall \candset \!\!\subseteq\! \Candidates, \forall \cand \!\in\! \Candidates\!: \mbm{\!\candset \!\cup\! \{\cand\}}{\Slots}{\Rel} \ge \mbm{\!\candset}{\Slots}{\Rel}
    \label{eqn:monotonicity}
\end{equation*}
and also submodular in $\candset$, which means that $\forall \candset \subseteq \Candidates, \forall \cand,\cand' \in \Candidates$ 
\begin{align*}
    \mbm{\candset \cup \{\cand\}}{\Slots}{\Rel} - \mbm{\candset}{\Slots}{\Rel} 
    \ge \mbm{\candset \cup \{\cand,\cand'\}}{\Slots}{\Rel} - \mbm{\candset \cup \{\cand'\}}{\Slots}{\Rel}.
    \label{eqn:submodularity}
\end{align*}
\end{lemma}

\begin{proof}
    Throughout, by a matching $\mathcal{M}$ from $\candset$ to $\mathcal{S}$, we refer to a set of candidate-slot pairs $\mathcal{M} \subset \candset \times \mathcal{S}$ such that all pairs are relevant ($R_{cs} = 1$ for all $(c, s) \in \mathcal{M}$) and no candidate or slot appears in more than one pair in $\mathcal{M}$. We call a matching $\mathcal{M}$ from $\candset$ to $\mathcal{S}$ maximum if $|\mathcal{M}|$ is maximized over all matchings from $\candset$ to $\mathcal{S}$. 
    
    For monotonicity, we know that every single matching from $\candset$ to $\Slots$ is also a possible matching from $\candset \cup \{\cand\}$ to $\Slots$, thus the size of the maximum matching from $\candset \cup \{\cand\}$ to $\Slots$ must be at least the size of the maximum matching from $\candset$ to $\Slots$. So Equation~\ref{eqn:monotonicity} holds.

    For submodularity, if we denote 
    \begin{align}
        M_0 &= \mbm{\candset}{\Slots}{\Rel} \\
        M_1 &= \mbm{\candset \cup \{\cand\}}{\Slots}{\Rel} \\
        M_2 &= \mbm{\candset \cup \{\cand'\}}{\Slots}{\Rel} \\
        M_{12} &= \mbm{\candset \cup \{\cand,\cand'\}}{\Slots}{\Rel}
    \end{align}

    We want to show that 
    \begin{equation}
         M_1 - M_0 \ge M_{12} - M_2.
        \label{eqn:submodularity-simple}
    \end{equation}

    Let us denote $M_1 - M_0$ as the LHS and $M_{12} - M_2$ as the RHS for Equation~\eqref{eqn:submodularity-simple}. Equation~\eqref{eqn:monotonicity} says both the LHS and RHS are nonnegative. Additionally, the LHS is upper bounded by $1$, as given any matching $\mathcal{M}$ of size $M_1$ from $X \cup \{C\}$ to $\mathcal{S}$, we can construct a matching $\mathcal{M}'$ from $X$ to $\mathcal{S}$ of size at least $M_1-1$ by removing the unique pair involving $C$ from $\mathcal{M}$, if it exists, which shows $M_0 \geq |\mathcal{M}'| \geq M_1 - 1$. By the same reasoning, the RHS is upper bounded by $1$, and indeed we also have $0 \leq M_2 - M_0 \leq 1$. Then since both the LHS and RHS can only take on values $0$ or $1$, in order to prove equation \eqref{eqn:submodularity-simple} it suffices to show that if the RHS equals $1$, then the LHS equals $1$.
    
    Assume the RHS equals $1$. Then any maximum matching $\mathcal{M}$ from $X \cup \{C,C'\}$ to $\mathcal{S}$ must include a pair $(C, S)$ for some $S \in \mathcal{S}$, or else $\mathcal{M}$ would also be a valid matching from $X \cup \{C'\}$ to $\mathcal{S}$, which would imply that $M_2 \geq |\mathcal{M}| = M_{12}$, violating the assumption. We split on two cases, which as we noted earlier are exhaustive.
    
    \emph{Case}: $M_2 - M_0 = 0$. Let $\mathcal{M}_0$ be some maximum matching from $X$ to $\mathcal{S}$. Since $M_2 = M_0$ by assumption, $\mathcal{M}_0$ is also a maximum matching from $X \cup \{C'\}$ to $\mathcal{S}$. Indeed, $\mathcal{M}_0$ is also a matching from $X \cup \{C, C'\}$ to $\mathcal{S}$, but not a maximum such matching, as $M_{12} > M_2$ by assumption. Therefore, by Berge's theorem, there must exist an ``augmenting path'' $P$ consisting of candidate-slot pairs $(c_1, s_1), \ldots, (c_T, s_T) \in (X \cup \{C, C'\}) \times \mathcal{S}$ for all $t$ such that all pairs are relevant, i.e. $R_{c_t, s_t} = 1$ for all $t$, and the path starts and ends on unmatched edges and alternates between matched and unmatched edges, i.e. $(c_t, s_t) \notin \mathcal{M}_0$ for all $t$ and $(c_t, s_{t-1}) \in \mathcal{M}_0$ for all $t \in \{2,\ldots,T\}$. But then since $P$ contains one more unmatched than matched edge, ``applying'' $P$ to $\mathcal{M}_0$ via the set difference operation gives a maximum matching from $X \cup \{C, C'\}$ to $\mathcal{S}$, since $\mathcal{M} := (\mathcal{M}_0 \setminus P) \cup (P \setminus \mathcal{M}_0)$ satisfies $|\mathcal{M}| = M_0 + 1 = M_2 + 1 = M_{12}$. But since $\mathcal{M}$ is a maximum matching from $X \cup \{C, C'\}$ to $\mathcal{S}$, we must have that $\mathcal{M}$ contains candidate $C$ as shown earlier, thus it cannot contain $C'$ by the alternating edges property of $P$ as both $C$ and $C'$ are unmatched in $\mathcal{M}_0$. But then $\mathcal{M}$ is a valid matching from $X \cup \{C\}$ to $\mathcal{S}$, hence $M_1 \geq |\mathcal{M}| = M_{12} = M_0 + 1$ as desired.
    

    \emph{Case}: $M_2 - M_0 = 1$. Let $\mathcal{M}$ be a maximum matching from $X \cup \{C,C'\}$ to $\mathcal{S}$; then we must have $|\mathcal{M}| = M_{12} = M_0 + 2$ by assumption. If candidate $C'$ does not appear in $\mathcal{M}$, then $\mathcal{M}$ is a valid matching from $X \cup \{C\}$ to $\mathcal{S}$, so $M_1 \geq |\mathcal{M}| = M_2 + 1 \geq M_0 + 1$ and we are done. Otherwise, assume $\mathcal{M}$ includes a pair $(C', S')$ for some $S' \in \mathcal{S}$. 
    But then $\mathcal{M}'' := \mathcal{M} \setminus \{ (C', S') \}$ is a valid matching from $X \cup \{C\}$ to $\mathcal{S}$ with $|\mathcal{M}''| = M_{12} - 1 = M_0 + 1$, hence $M_1 \geq M_0 + 1$ as desired. 
\end{proof}


The monotone submodularity of $\mbm{\candset}{\Slots}{\Rel}$ implies that our estimate $\expmatchest(\candset)$ is also monotone submodular, since this property is closed under addition.

\begin{lemma} \label{lem:submodmhat}
    For any sample of relevance matrices $\RelSamples= [\Rel_1,...,\Rel_\n]$ and set of slots $\Slots$, $\expmatchest(\candset)$ from Equation~\eqref{eq:expmatchest} is monotone in $\candset$
\begin{equation*}
    \forall \candset \subseteq \Candidates, \forall \cand \in \Candidates: \expmatchest(\candset \cup \{\cand\}) \ge \expmatchest(\candset)
\end{equation*}
and also submodular in $\candset$, which means that $\forall \candset \subseteq \Candidates, \forall \cand,\cand' \in \Candidates$ 
\begin{eqnarray*}
    \expmatchest(\candset \!\cup\! \{\cand\}) \!-\! \expmatchest(\candset) 
    \ge \expmatchest(\candset \!\cup\! \{\cand,\cand'\}) \!-\! \expmatchest(\candset \!\cup\! \{\cand'\}).
\end{eqnarray*}
\end{lemma}

\begin{proof}
    We know from Lemma~\ref{lem:submodmbm} that each each $\mbm{\candset}{\Slots}{\Rel_i}$ is monotone and submodular. This means that $\expmatchest(\candset)$ is a sum of monotone submodular function, and it is well known that sum is monotone submodular as well.
\end{proof}


We are now in a position to complete the proof of Theorem~\ref{thm:exp}.

\begin{proof}[Proof of \theo{thm:exp}]
    Our goal is to bound with high probability the suboptimality $M(X_k) - M(X_k^*)$ between the candidate set $X_k$ output by the model and the optimal candidate set $X_k^*$.

    First, note that the approximation guarantee of the greedy algorithm for monotone submodular maximization with a cardinality constraint \cite{nemhauser1978analysis} guarantees
    \begin{align}
        \hat{M}(X_k) \geq (1-1/e)\hat{M}(\hat{X}_k^*) \geq (1-1/e)\hat{M}(X_k^*), \label{eq:greedyempirical}
    \end{align}
    where $\hat{X}_k^*=\argmax_{|X|=k} \hat{M}(X)$ is the optimal set on the Monte-Carlo samples. The second inequality follows from $\hat{M}(X_k^*) \le \hat{M}(\hat{X}_k^*)$.

    To get a bound in terms of $M(.)$ instead of $\hat{M}(.)$, we need to bound the error due to Monte-Carlo sampling. We start with the following equivalent expansion of Equation~\eqref{eq:greedyempirical}.
    \begin{align*}
        M(X_k) \ge\left(1\!-\!\frac{1}{e}\right) \left(M(X_k^*) 
        \left(\hat{M}(X_k^*) \!-\! M(X_k^*)\!\right) \right) + \left(M(X_k) \!-\! \hat{M}(X_k)\!\right) 
    \end{align*}
    First, we upper bound $\hat{M}(X_k^*)-M(X_k^*)$, for which we can use Hoeffding's inequality since for any $X$ it holds that $E_R[\hat{M}(X)]=M(X)$ and the $MBM(X,\mathcal{S}|R_i)$ ($i \in \{1, \ldots, n\}$) are i.i.d.\ Monte-Carlo samples with $0 \leq MBM(X,\mathcal{S}|R_i) \leq s$.
    \begin{align*}
        P(M(X) - \hat{M}(X) > \epsilon) \leq \exp ( -2n\epsilon^2/s^2 ).
    \end{align*}
   We thus get for our particular $X_k^*$ that with probability $0 \le \delta_1 \le 1/2$
   \begin{eqnarray}
       \hat{M}(X_k^*) - M(X_k^*) \le s\sqrt{\frac{\ln(1/\delta_1)}{2n}} := \epsilon_1. \nonumber
   \end{eqnarray}

   Second, we need to upper bound $M(X_k)-\hat{M}(X_k)$. 
   Since $X_k$ is selected on the same Monte-Carlo sample we evaluate it on, we need to ensure uniform convergence over all $X$. We thus take the union bound over the set $\mathcal{H}_k$ of all possible candidate sets of size $k$:
    \begin{align*}
        P(\max_X ( M(X) - \hat{M}(X) ) > \epsilon) &\leq |\mathcal{H}_k| \exp ( -2n\epsilon^2/s^2 ) \\
        &\leq \left(\sqrt{2\pi k} \left(\frac{k}{e}\right)^k e^{\frac{1}{12k}}\right) \exp \left( \frac{-2n\epsilon^2}{s^2} \right)
    \end{align*}
    The second step uses Stirling's inequality, since $|\mathcal{H}_k| = k!$. By letting the final expression equal $\delta_2$ and solving for $\epsilon$, we get that with probability $0 \le \delta_2 \le 1/2$ it holds for all $X$ (and therefore also for any $X_k$ our algorithm picks) that
\begin{eqnarray}
     & M(X_k) - \hat{M}(X_k) \le s\sqrt{\frac{(k\ln k - k + O(\ln k)) + \ln(1/\delta_2)}{2n}} := \epsilon_2. \nonumber
\end{eqnarray}
    
    Putting these bounds together, we get that for all $0 < \delta_1, \delta_2 < 1/2$, we have with probability $1 - \delta_1 - \delta_2$,
\begin{align*}
    M(X_k) &\ge \left(1-\frac{1}{e}\right) (M(X_k^*) - \epsilon_1) - \epsilon_2  \\
    &\ge \left(1-\frac{1}{e}\right)M(X_k^*) - (\epsilon_1 + \epsilon_2).
\end{align*}

Setting $\delta_1 = \delta_2 = \delta/2$ gives the claimed bound.
\end{proof}

\subsection{Computational Efficiency of \algname and Improvements} \label{sec:efficiency}

The \algname algorithm as written in \alg{alg:greedy} is optimized for clarity, but there are a number of improvements that can substantially speed up computation. To motivate these improvements, we first analyze the runtime complexity of \alg{alg:greedy}.

For computing the top $k$ positions of the ranking when there are $\CandidatesNum$ candidates, $\SlotsNum$ slots, and $\n$ Monte-Carlo samples, the greedy maximizer inside \alg{alg:greedy} will evaluate $O(k \, \CandidatesNum)$ sets. For each such set, it will find the MBM solutions of all bipartite graphs from $\RelSamples$, which takes $\n \, O(\CandidatesNum \, \SlotsNum \, \sqrt{\CandidatesNum + \SlotsNum})$ time per set when using the classic Hopcroft-Karp MBM algorithm \cite{hopcroft1973n}. So, a naive implementation will take $O(k \, \n \, \CandidatesNum^2 \, \SlotsNum \, \sqrt{\CandidatesNum + \SlotsNum})$ time. However, this implementation is unnecessarily slow.

We can improve the time efficiency of \textit{MatchRank} by following the principle that any unmatched candidate can increase the matching size by at most 1. So, if we are given a candidate set $\candset \subseteq \Candidates$ and an unmatched candidate $\cand$, finding the $\mbm{\candset \cup \{\cand \}}{\Slots}{\Rel}$ can be reduced to determining if there is an augmenting path to the matching of $\mbm{\candset}{\Slots}{\Rel}$ starting from $\cand$. If such an augmenting path exists, then we can extend the matching and $\mbm{\candset \cup \{\cand \}}{\Slots}{\Rel} = \mbm{\candset}{\Slots}{\Rel} + 1$. It no augmenting path exists, then we will know that $\mbm{\candset \cup \{\cand \}}{\Slots}{\Rel}=\mbm{\candset}{\Slots}{\Rel}$ and the matching remains unchanged. Therefore, it will take $O(\CandidatesNum \, \SlotsNum)$ time (a BFS) per each unmatched candidate per ranking step, and we obtain an $O(k \, \n \, \CandidatesNum \, (\CandidatesNum \, \SlotsNum)) = O(k \, \n \, \CandidatesNum^2 \, \SlotsNum)$ algorithm.  

If we consider the typical scenario where $\CandidatesNum > \SlotsNum$, we can improve the time complexity by keeping a list of unmatched slots instead of candidates for each bipartite graph, and on each ranking step we start from each unmatched slot and follow the BFS to find all augmenting paths that end on unmatched candidates. We require $O(\CandidatesNum \, \SlotsNum^2)$ time per ranking step, and in total only $O(k \, \n \, \CandidatesNum \, \SlotsNum^2)$ time. We also note that on each ranking step each Monte-Carlo estimate $\Rel_i$ is independent and with perfect parallelism, we could eliminate the dependency over $n$ to get $O(k \, c \, s^2)$ time complexity.




A further improvement in runtime can be achieved by exploiting that \algname is a greedy algorithm for maximizing a submodular objective. For any such algorithm, we can use lazy evaluation \cite{minoux1978accelerated} to accelerate the ranking process in practice\footnote{We run all of our experiments with lazy greedy.}. Lazy greedy maintains a priority queue of stale marginal gains to reduce unnecessary computation of marginal gains for many examples per step. Since marginal gains can never increase due to the submodularity of the objective, the stale marginal gains provide an upper bound on the improvement. Thus, if a stale marginal gain is not large enough to propose a candidate as a greedy maximizer, then recomputing its marginal gain is not necessary. This is particularly effective in our matching scenarios, since many candidates are not relevant for some slots. This means they will have small marginal gains even in the first step of ranking, and we can significantly reduce computation of their marginal gains during subsequent iterations when using lazy greedy.

Finally, we could replace the greedy algorithm with some approximate version of greedy that is substantially faster, such as stochastic greedy \cite{mirzasoleiman2015lazier} and threshold greedy \cite{badanidiyuru2014fast} for monotone submodular maximization problem. We have performed initial evaluations of these methods with promising results, but did not find a need for them for our experiments. Generally, we found the exact greedy algorithm to be tractable for datasets with up to 50,000 candidates. 

\begin{figure}[t]
    \centering
    \includegraphics[width=0.65\columnwidth]{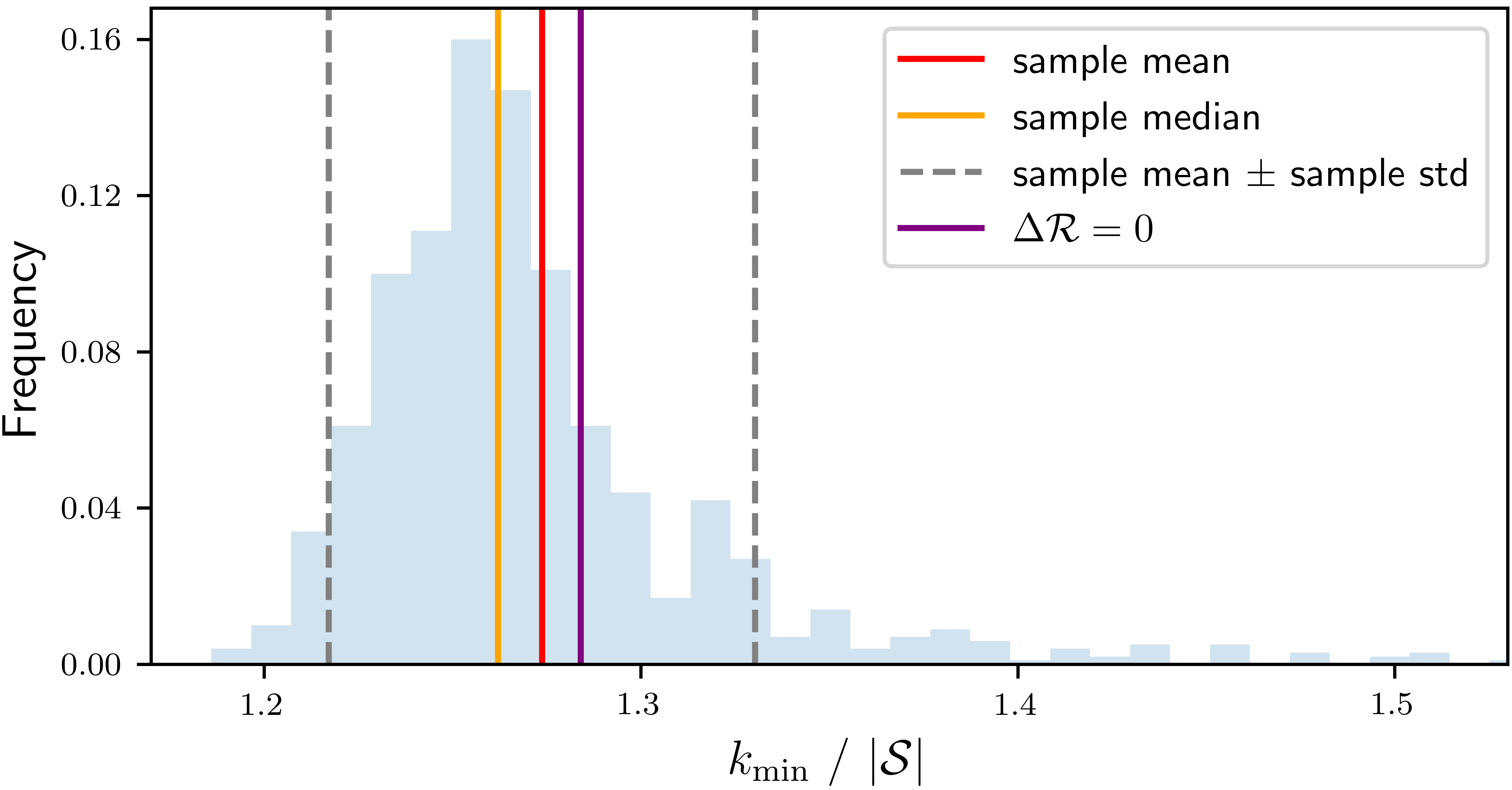}
    \vspace*{-0.3cm}
    \caption{Synthetic Datasets: Histogram of $k_{\min} / | \Slots|$ for \algname under default settings for multipe draws of the ground-truth relevances from $P(\Rel)$. The vertical lines are the mean, median, and standard deviation of the histogram. $\Delta \RelSamples = 0$ is the position in the ranking when all slots in the Monte-Carlo $\RelSamples$ samples are filled and the marginal gain of adding a another candidate becomes 0. \label{fig:synthetic-matchrank}}
\end{figure}

\subsection{Variability \& robustness of \algnamens}

To further understand how the Monte-Carlo samples affect the accuracy of \algnamens, Figure~\ref{fig:synthetic-matchrank} shows a histogram of $k_{\min} / |\Slots|$ for \algname with $n=200$ Monte-Carlo samples when evaluating the ranking on true relevance matrices sampled from $P(\Rel)$. The distribution has heavier tails on the right, and the mean is higher than the median. Most interestingly, however, is the location of the line where the ranking matches all slots on all 200 Monte-Carlo samples (denoted with $\Delta\RelSamples = 0$). Even though \algname perfectly matches the Monte-Carlo samples, the histogram shows that there are still many other "test" samples (specifically, 24.9\% of the samples) from $P(\Rel)$ that have unmatched slots. This is a form of overfitting that a larger Monte-Carlo sample can avoid, and thus provide improved performance for \algnamens.

\paragraph{Analyzing the variability of \algnamens}

We want to understand the variability of $k_{\min} / |\Slots|$ of \algname across ground truth relevance matrices. So we prepare a histogram of $k_{\min} / |\Slots|$ for \algname evaluating on $\RelSamples_\text{test}$ in default settings as shown in Figure~\ref{fig:synthetic-matchrank}. This figure demonstrates that the distribution of $k_{\min} / |\Slots|$ is slightly right-tailed as the sample mean is higher than the sample median. This could be explained as \algname is prone to overestimating candidates' probability of relevances due to the nature of greedy. Therefore, it is highly probable that some slots are not filled by relevant candidates yet even when \algname believes all slots have been filled. Such phenomenon is illustrated by the fact that there are still 24.9\% $\Rel$ samples for which slots have not been filled after $\Delta \RelSamples$ becomes 0 (the purple line) in the Figure~\ref{fig:synthetic-matchrank}.

\paragraph{Analyzing the Robustness of \algnamens: Effect of $\hat{P}(R)$ Misspecification} \label{para:synthetic-misspecification}

\begin{figure}[t]
    \centering
    \includegraphics[width=0.65\columnwidth]{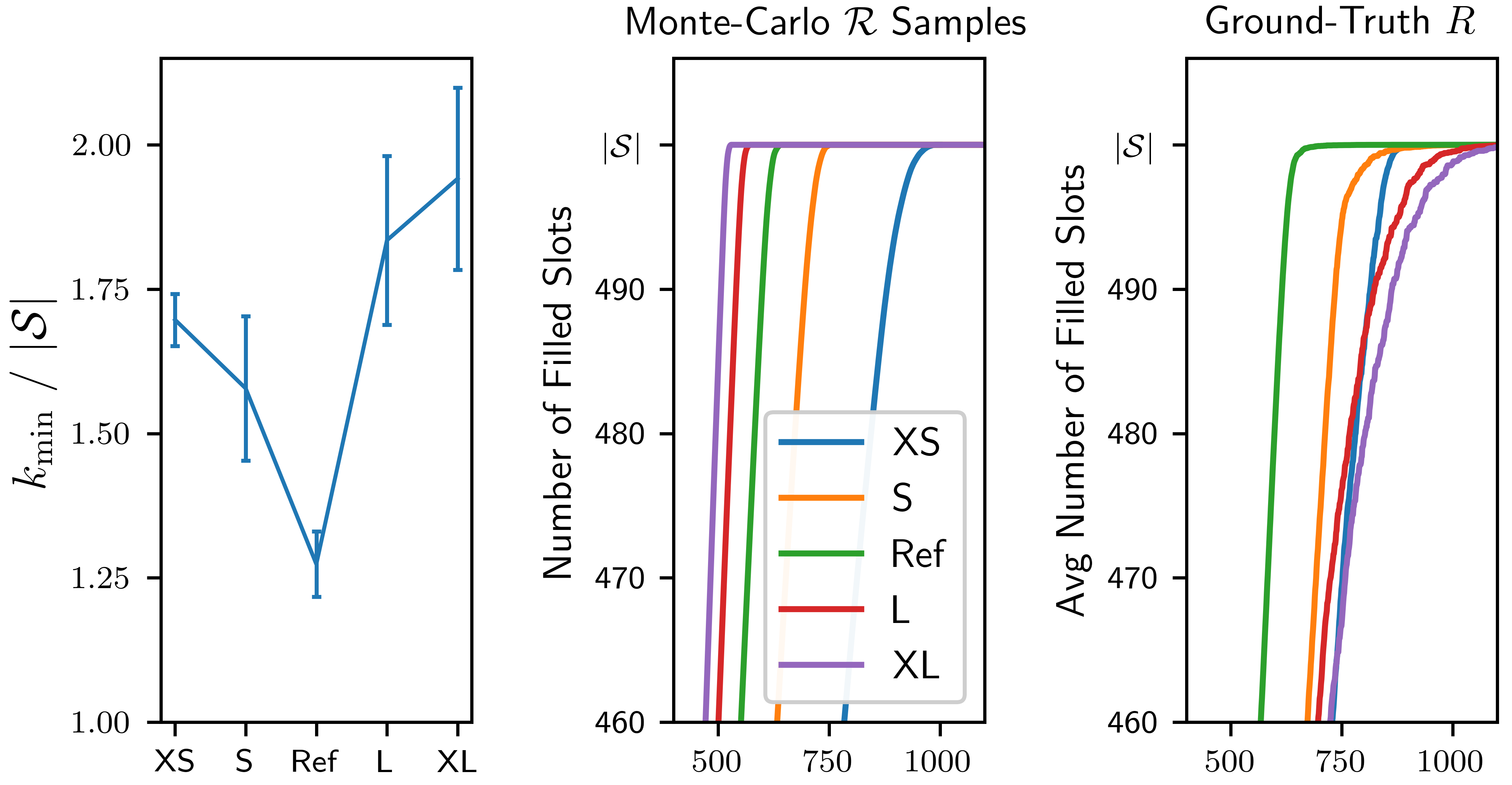}
    \vspace*{-0.3cm}
    \caption{Effect from $\PR$ misspecification, where Ref is the correct model. The left subfigure shows mean and standard deviation of $k_{\min} / |\Slots|$. \label{fig:ou}}
\end{figure} 

We now investigate how robust \algname is against an inaccurately learned $\hat{P}(R)$. We draw the ground-truth relevance labels from a model $P(R)$ with $p_\text{base}=0.3$ (Ref), but draw the Monte-Carlo samples from misspecified models $\hat{P}(R)$ with $p_\text{base}$ set to 0.1 (XS), 0.2 (S), 0.4 (L), and 0.5 (XL). Figure~\ref{fig:ou} shows that \algname performs best for the correctly specified $\hat{P}(R)$ as expected. Among the misspecified $\hat{P}(R)$, we can observe that \textit{MatchRank} is more vulnerable to overestimation of relevance probabilities than underestimation. This can be explained as follows. If \algname is erroneously convinced that a slot is filled with high probability, it will not add an alternative candidate for this slot to the ranking. This fact is illustrated in the middle subfigure of Figure~\ref{fig:ou}, where \algname fills all slots in the Monte-Carlo $\RelSamples$ samples faster as the overall relevance level increases. However, the average number of filled slots using the ground-truth relevances $\Rel$ is not necessarily higher, and in fact both L and XL have fewer filled slots per ranking step than S and XS for $\Rel$. So it may be advisable to clip $\hat{P}(R)$ to some maximum value that is well below 1 to increase robustness to misspecification.

\end{document}